\documentclass[11pt]{article}%
\usepackage[latin1]{inputenc}   
\usepackage[italian]{babel}     
\font\ninerm=cmr9

\def\ket#1{\vert\, #1\,\rangle}

\title{\bf\Large On the measurement problem in quantum mechanics: a simple proposal}
\author{Luigi E. Picasso
\\
\small{Dipartimento di Fisica, Universit\`a di Pisa, Italy}}
\date{}
\begin{document}
\maketitle
\begin{center}
{\bf Abstract}
\end{center}
\noindent Some of the problems connected with the interpretation of
quantum mechanics are enumerated, in particular those related to
some well known paradoxes and, above all, to the measurement
process. We then show how the so called ``\emph{Laboratory Physics
Assumption}'' introduced in \cite{Picasso}, which considers as
``observables'' only the self-adjoint operators corresponding to
existing measuring instruments,  can propose a new perspective on
the aforementioned problems and can replace the wavefunction
collapse postulate.
\par\bigskip
\noindent
{\bf keywords}\quad {\ninerm Quantum mechanics $\cdot$ Pure and mixed states $\cdot$ Wavefunction collapse}

\section{Introduction}
\noindent Quantum mechanics is the theory of atoms and molecules: it
explains why atoms absorb and emit light of definite frequencies,
how atoms can combine to generate molecules. For instance, it
explains not only how hydrogen and oxygen atoms can combine to form
water, but also the shape of its molecules. Moreover, since matter
is made of atoms and molecules,  we can say that every object is as
such thanks to the rules of quantum mechanics.

The peculiarity of quantum mechanics is that, in contrast to the
classical theory, it is not  deterministic: given the state of a
system, {\it i.e.}, knowing how it has been prepared, though it is
true that -- thanks to the Schr\"odinger equation -- its evolution
at any later time is well determined, it is not possible, in
general, to predict the value of the physical quantities, such as
energy, momentum, position etc., related to the system.  If one
wants to know the value  of a specific observable, the system must
be subjected to a measurement process and only the probabilities of
the results of such measurements can be predicted.
It is clear therefore, that there is a twofold evolution underlying
any quantum system. The first one is governed by the Schr\"odinger
equation and is strictly deterministic; the second one reveals
itself when a measurement is performed: the state of the system
collapses, with a given probability $p_i$, into one of the
eigenstates $\ket{\xi_i}$ of the measured observable $\xi$.

Ever since the work of Born \cite{Born}, which constituted a
milestone in the interpretation of quantum mechanics, this situation
has been considered a weak point of the theory. Indeed, the
measuring apparatus must be a classical object, {\it i.e.}, not
subject to the laws of quantum mechanics: its `pointers' are
classical because they must be recordable by the observer.
This led to the conclusion that there is some limit beyond which
quantum mechanics must give way to classical physics: Bohr's opinion
was that there must be a dividing line, not necessarily a fixed one,
between the quantum and classical domains.

At this point, however, Griffiths' \cite{Griffiths}  words sound
really appropriate: ``\dots after all, what is special about a
quantum measurement? All real measurement apparatus is constructed
out of aggregates of particles to which the laws of quantum
mechanics apply, so the apparatus ought to be described by those
laws, and not used to provide an excuse for their breakdown.''
The natural consequence of Griffiths' words is that the so called
measuring apparatus must be included as part of the system, giving
rise to a Grand-System. Nonetheless, this idea is not free from
difficulties either: for instance, what operates the measurements on
the Grand-System? This is an endless question that led Wigner to
suggest that the wavefunction collapse is ultimately due to our
consciousness, even if later he distanced himself from this
idea \cite{Ballentine}.

This article stems from two articles, \cite{Picasso} and
\cite{BracciPicasso}, and is justified by the need for merging and
synthesizing the results in the mentioned papers and, above all, to
present their contributions in a more systematic way. This paper is
organized as follows: in the next section the problems of the
measurement in quantum mechanics are summarized, then some of the
paradoxes, as the ``de Broglie box'' and the ``Schr\"odinger cat'',
are considered. In Section 3 the so called \emph{Laboratory Physics
Assumption} (LPA) is introduced and its consequences are discussed
in Section 4, where it is shown how the aforementioned problems can
be considered under a new perspective and the wavefunction collapse
postulate can be replaced by the LPA. In Section~5 the main points
of this article are highlighted.

\section{Interpretative challenges of quantum mechanics}

\subsection{The measurement problem}

While the wavefunction collapse postulate had a place within a
theory where there is a clear distinction between the quantum and
the classical domains, it is now difficult, with the Grand-System,
to pinpoint how it can be replaced.
Indeed, let  the state of the quantum system be $\sum\nolimits_i
a_i\ket{\xi_i}$, $\mathbf\Xi$ the apparatus which measures the
observable $\xi$ and $\,\ket{\Xi_i}$ the ``pointer states'', {\it
i.e.}, the mutually orthogonal states of the apparatus (detector)
when the display exhibits the value $\xi_i\,$; the pointer states
generate the Hilbert space ${\cal H}_{\rm D}$ of the detector. For
the Grand-System, prior to the interaction between the quantum
system and the measuring device, the state is the product
\begin{equation}
\sum\nolimits_i a_i\ket{\xi_i}\ket{\Xi_0} \label{GS-in}
\end{equation}
where $\ket{\Xi_0}$ is the state of the untriggered measuring
device. The final state is
\begin{equation}
U(t_{\rm fin})\,\sum\nolimits_i a_i\ket{\xi_i}\ket{\Xi_0} =
\sum\nolimits_i a_i\ket{\xi_i}\ket{\Xi_i}. \label{GS-fin}
\end{equation}

This should be a pure state of the Grand-System but it is not clear
at which point the measurement takes place.

In any case, since a set of $\xi_i$ values, each with frequency $p_i
= \vert a_i\vert^2$, are somehow registered  in correspondence with
the states $\ket{\xi_i}\ket{\Xi_i}$ of the Grand-System, then the
Grand-System is only formally described by Eq.\,(\ref{GS-fin}) but
it actually behaves as the statistical mixture
\begin{equation}
\big\{\ket{\xi_i}\ket{\Xi_i},\vert
a_i\vert^2;\;i=1,\cdots\big\}\label{grand-mixture}.
\end{equation}

However, this result is clearly incompatible with the rules of
quantum mechanics: the deterministic unitary evolution governed by
Schr\"o\-dinger's equation cannot transform pure states into
statistical mixtures.

\subsection{The paradoxes of quantum mechanics}

The uneasiness produced by the interpretation of quantum mechanics
is witnessed by the many paradoxes that were put forward in the
early years of quantum mechanics.
The most popular ones are the ``de Broglie box'' and
``Schr\"odinger's cat''. Both have in common the problems raised
by the superposition principle when applied at the macroscopic
scale. We shall eventually see how they can be reconciled with our
common sense.

Let us begin with de Broglie's box. An electron is in a stationary
state within a box. By taking all necessary precautions, the box is
split into two equal boxes by means of a sliding diaphragm, then the
two boxes -- let us say the right one and
the left one -- are taken far apart from each other.%
\footnote{In a more realistic version we could consider, for
instance, a neutron after passing through a beam-splitter as the
ones used in the Bonse-Hart neutron spectrometer.}
According to quantum mechanics, the electron is in a state $\ket
A\,$ that is a superposition of the state $\ket r\,$, corresponding
to the electron being in the right box, and of the state $\ket l\,$,
in which the electron is in the left box. Since one only knows that
the probabilities of finding the electron in any one of the boxes is
1/2, the state $\ket A$ can be expressed as
\begin{equation}
\ket A = \frac{1}{\sqrt2}\big(\ket r +e^{\rm i\,\phi}\,\ket l\big)
\; . \label{(1)}
\end{equation}

How can the phase $\phi$ be determined? This is possible only if one
can measure observables that have nonvanishing matrix elements
between $\ket r\,$ and $\ket l\,$, since only in that case the two
states can be made to interfere. Certainly, there are infinitely
many self-adjoint operators in the Hilbert space of the system
having this property, but are there actually any \emph{real}
measurement instruments corresponding to those operators? If the two
boxes are sufficiently far apart, we are sure that in no laboratory
such instruments do exist. All~the same, according to the principles
of quantum mechanics, the state of the system is a pure state, even
if in practice it behaves as the statistical mixture
\begin{equation}
\big\{\ket r,\textstyle{\frac{1}{2}}\,;\ \ket
l,{\frac{1}{2}}\big\}.\label{(2)}
\end{equation}

Also in the case of Schr\"odinger's cat paradox we are sure that
there do not exist observables ({\it i.e.}, measurement instruments)
with nonvanishing matrix elements between the dead and the alive
cat: indeed, if such instruments existed, then with a couple of
measurements it would be possible, with  nonvanishing probability,
to resuscitate a dead cat. Therefore, in this case too, the state of
the system (cat + the ampoule of cyanide), although pure, behaves as
a statistical mixture.

As already noticed, the paradoxical aspects of the above examples
stem from the superposition principle being applied to
macroscopically distant states in the first case, and to a
macroscopic object as the cat in the second case.

Should we then accept that quantum mechanics applies only to the
microscopic world? Certainly not, since we know several cases of
macroscopic systems that exhibit quantum properties, such as helium
at very low temperatures or the superconductors. Therefore, given
that with Griffiths quantum mechanics applies to every system, how
can we know if a given system can be described and treated with the
laws of classical physics?

It is widely accepted that the postulates, or rather the
interpretation, of quantum mechanics need some revision. In this
regard, interesting discussions of the various problems and
proposals that have been put forward can be found both in the book
of John Bell ``Speakable and Unspeakable in Quantum Mechanics''
\cite{Bell}, especially in Chapter~23 (``Against «measurement»''),
and, more recently, in \cite{Sofia}. The LPA proposal discussed in
the subsequent section offers an original contribution to the
subject that allows one to deal with most of the critical issues
raised in the literature in a straightforward way, without resorting
to hypotheses that may sound rather artificial or in contrast to the
commonly agreed principles of quantum mechanics.

\section{LPA}
In the papers \cite{Picasso,BracciPicasso}
a proposal called \emph{Laboratory Physics Assumption} (LPA) was put
forward as a substitute for the wavefunction collapse postulate. It
starts from the observation
that in quantum mechanics it is
assumed that to every self-adjoint operator $\xi$ in the Hilbert
space of any system there corresponds an instrument $\mathbf\Xi$
which measures the observable $\xi$.
The LPA considers
``observables'' only the self-adjoint operators corresponding to
existing measuring instruments in {\sl some}
laboratory in the world.%
\footnote{This requirement may seem vague if not unviable: we shall
see, however, that no problem arises in any of the situations in
which it will be applied.}

This assumption has strong effects on the very concept of ``state'':
as it has been shown in the examples of the previous section,
according to the classical rules of quantum mechanics pure states
may behave as statistical mixtures. The LPA asserts that in all such
situations, {\it i.e.}, in the absence of observables with non
vanishing matrix elements between the various component of the
mixture, those states \emph{are} actually statistical mixtures.
Thus, for instance, in the case of the de Broglie box the electron
is, prior to observation, either in the $\ket r$ \emph{or} in the
$\ket l$ state, each with probability $1/2$, and Schr\"odinger's cat
is either dead \emph{or} alive even before opening the box.

In other words, the superposition principle can be applied only to
those states to which it makes sense, {\it i.e.}, to states such
that instruments capable of measuring the relative phases do exist.
Thus, even in the Wigner's friend paradox -- where Wigner's friend
belongs to the laboratory where the box with the cat and the ampoule
of cyanide are present -- it does not make sense for Wigner, while
outside the laboratory, to think of a coherent superposition between
$\ket{cat\; alive, friend\; happy}$ and $\ket{cat\; dead, friend\;
sad}$: a relative phase factor could not be measured.

Therefore, the properties (``pure'' or ``not pure'') of a state are
not intrinsic but depend on which observables ({\it i.e.},
measurement instruments) do exist today. Indeed, we can say that,
while according to the rules of quantum mechanics a dead cat can
\emph{in principle} be resuscitated, according to the LPA
\emph{today} a dead cat cannot  be brought to live again. We have
emphasized the word \emph{today} because we cannot exclude that the
observables that are not available today may be available in the
future and
states that today are statistical mixtures tomorrow will be pure.%
\footnote{The literature already offers examples of this kind: a few
decades ago the two photons emitted in the cascade
$^1S_0\to\,^1P_1\to\,^1S_0$ of the Calcium atom used by A.~Aspect
and collaborators to prove the violation of Bell's inequalities,
could be considered a pure state only for distances between the two
photons of a few tens of meters, today the experiment by Yin et al.
\cite{Jin} shows that coherence can be maintained for more than one
thousand kilometers.}

\section{Consequences}
As for the paradoxes of quantum mechanics, the LPA allows one to be
freed from the embarrassing necessity of applying the superposition
principle in situations where its application is indeed unfitting.
However, the relevant consequence of the LPA is concerned with the
measurement problem.

The first unavoidable question is about the status of a measurement
instrument: with Griffiths, we take for granted that a measurement
instrument is  a quantum system. It is quite obvious that the only
pure states of the instrument are the pointer states: this means
that coherent superpositions of the pointer states will never be
observed as the result of a measurement. According to the LPA, this
is equivalent to stating the non-existence in ${\cal H}_{\rm D}$ of
observables with nonvanishing matrix elements between the pointer
states, and this is taken as an inalienable part of the definition
of a measurement instrument. Therefore not only all the states as in
Eq.\,(\ref{GS-fin}) \emph{behave} as statistical mixtures, but they
\emph{are}, indeed. What was not possible according to the classical
rules of quantum mechanics, {\it i.e.}, the transition from pure
states to statistical mixtures, is now possible thanks to the LPA.

The difference between the LPA and the postulate of the wavefunction
collapse is only conceptual: while the second is an \emph{ad hoc}
postulate, the LPA reproduces the same results on the basis of a
precise physical fact (the absence of observables in ${\cal H}_{\rm
D}$ with nonvanshing matrix elements between the pointer states) and
is responsible for the transition from an initial pure state
as Eq.\,(\ref{GS-in}) to a statistical mixture as Eq.\,(\ref{grand-mixture}).%
\footnote{In \cite{BracciPicasso} a model is presented of a
hamiltonian $H$ that, with some approximations, is responsible for
the transition from Eq.\,(\ref{GS-in}) to Eq.\,(\ref{GS-fin}).}
Hence, it should be clear that, in the LPA perspective, the
phenomenon commonly referred to as the wavefunction collapse must be
ascribed to the measurement apparatus.

\section{Conclusions}
The main points of this article are

\noindent 1.\ the properties of the states (pure, not pure) are
determined only by the observables to which there correspond
existing measuring instruments;

\noindent 2.\ a different, but equivalent way of formulating the LPA
is the following: the superposition principle can only be applied to
states such that instruments capable of measuring the relative
phases do exist (thus, a coherent superposition principle can be
applied to particular states only). Specifically, the superposition
principle cannot be applied to the pointer states of a measuring
instrument;

\noindent 3.\ the state of the Grand-System, although starting as a
pure state and being its time evolution formally determined by a
unitary operator, ends up as a statistical mixture. At this point
the measurement has taken place, irrespective of whether observed or
not.

A point that in this paper has only been touched upon (but widely
discussed in\,\cite {Picasso}) is that with the LPA nothing is
definitive: since all depends on the actually existing observables,
what today is a statistical mixture, tomorrow may be a pure state; a
detector behaves as a classical object as long as no observable is
available (in ${\cal H}_{\rm D}$) which connects different pointer
states. We don't know how this perspective may be considered from a
philosophical viewpoint, anyhow it is worth noting that in the
measuring process the LPA reproduces the same results as the
wavefunction collapse postulate: what changes is only the
interpretation of the process.

\section*{Acknowledgements}
I gratefully acknowledge Luciano Bracci
for a critical reading of the manuscript and several useful
discussions.
\par\bigskip
The author has no conflicts to disclose.

\renewcommand{\refname}{References}


\begin{thebibliography}{99}

\bibitem{Picasso}Luigi E. Picasso, ``On the Concept of State in Quantum Mechanics:
Another Way to Decoherence?'' Int. J. Theor. Phys. \textbf{62} (2), (2023).

\bibitem{Born}Max Born (1926), ``Zur Quantenmechanik der Stoßvorgänge''
[On the quantum mechanics of collisions]. Zeitschrift für Physik. 37 (12): 863--867.

\bibitem{Griffiths}Robert B. Griffiths \textit{Consistent Quantum theory}, Cambridge University Press, 214 (2002).

\bibitem{Ballentine}Leslie E. Ballentine (2019), ``A Meeting with Wigner''. Foundations of Physics. 49 (8): 783--785.

\bibitem{BracciPicasso}Luciano Bracci and Luigi E. Picasso, ``About The Measurement Process in Quantum Mechanics''
Int. J. Theor. Phys. \textbf{64}, 220 (2025).
https://doi.org/10.1007/s10773-025-06091-6

\bibitem{Bell}John S. Bell and Alan Aspect, \textit{Speakable and Unspeakable in Quantum Mechanics: Collected
Papers on Quantum Philosophy}. 2nd ed. Cambridge University Press; 2004.

\bibitem{Sofia} Sofia D. Wechsler, ``The Quantum Mechanics {\sl Needs} the
Principle of Wave-Function Collapse--But This Principle Shouldn't Be
{\sl Misunderstood}\,''.
Journal of Quantum Information Science,
\textbf {11}, 42-63 (2021).

\bibitem{Jin}Yin et al.,
``Satellite-Based Entanglement Distribution Over 1200 kilometers,''
Science \textbf {356}, 1140-1144 (2017).
\end{thebibliography}
\end{document}